\documentclass{epl}
\usepackage{graphicx,amsmath,amssymb}
\usepackage{bm}

\title{Low frequency Rabi spectroscopy for a dissipative two-level system}
\shorttitle{Low frequency Rabi spectroscopy}
\author{Ya. S. Greenberg\inst{1}\thanks{E-mail: \email{greenbergy@online.nsk.su}} \and E.~Il'ichev\inst{2} \and A. Izmalkov\inst{2}}
\institute{
  \inst{1}Novosibirsk State Technical University, 20 K. Marx Ave., 630092 Novosibirsk, Russia\\
  \inst{2} Institute for Physical High Technology, P.O. Box 100239, D-07702 Jena, Germany
}


\begin{document}

\maketitle

\begin{abstract}

We have analyzed the interaction of a dissipative two level quantum system with high and low frequency excitation. The system is continuously
and simultaneously irradiated by these two waves. If the frequency of the first signal is close to the level separation the response of the
system exhibits  undamped low frequency oscillations whose amplitude has a clear resonance at the Rabi frequency with the width being dependent
on the damping rates of the system. The method can be useful for low frequency Rabi spectroscopy in various physical systems which are described
by a two level Hamiltonian, such as nuclei spins in NMR, double well quantum dots, superconducting flux and charge qubits, etc. As the examples,
the application of the method to a nuclear spin and to the readout of a flux qubit are briefly discussed.
\end{abstract}

It is well known that under resonant irradiation a quantum two
level system (TLS) can undergo coherent (Rabi) oscillations. The
frequency of these oscillations is proportional to the amplitude
of the resonant field~\cite{Rabi} and is much lower than the gap
frequency of TLS. The effect is widely used in molecular beam
spectroscopy ~\cite{beam}, and in quantum optics~\cite{Raimond}.

During the last several years it has been proven experimentally
that Rabi spectroscopy can serve as a valuable tool for the
determination of relaxation times in solid state quantum
mechanical two-level systems, qubits, to be used for quantum
information processing~\cite{Makhlin}. These systems normally are
strongly coupled to the environment, which results in the fast
damping of Rabi oscillations. It prevents the use of conventional
continuous measurements schemes for their detection, though the
special schemes for the detection of coherent oscillations through
a weak continuous measurement of a TLS were proposed
in~\cite{Averin, Korotkov1, Korotkov2}. That is why Rabi
oscillations are measured with the pulse technique through the
statistic of switching events of the occupation probability
between two energy levels with excitation and read out being taken
at the gap frequency of TLS, which normally, lies in GHz
range~\cite{Nakamura, Vion, Martinis, Chiorescu}. The main
drawback of this technique is that it requires sophisticated high
frequency readout electronics.

In this work we propose a new experimental method for a Rabi
spectroscopy of a TLS, when readout electronics is continuously
swept across the low (compare to the gap) Rabi resonance. The
method is rather general and can be applied to a great variety of
two level systems.

We consider a TLS which is irradiated continuously by two external
sources. The first with a frequency $\omega_0$, which is close to
the energy gap between the two levels, excites the low frequency
Rabi oscillations. Normally, Rabi oscillations are damped out with
a rate, which is dependent on how strongly the system is coupled
to the environment. However, if a second low frequency source is
applied simultaneously to TLS it responds with persistent low
frequency oscillations. The amplitude of these low frequency
oscillations has a resonance at the Rabi frequency with the width
being dependent on the damping rates of the system. Note, that
this approach has a well known classical analog. Indeed, a damping
rate of a classical oscillator can be easily obtained from its
amplitude-frequency characteristics.

We start with a Hamiltonian of a driven TLS , which is subjected to both high and low frequency excitation:
\begin{equation}\label{Ham1}
H = \frac{\Delta }{2}\sigma _x  + \frac{\varepsilon }{2}\sigma _z
- \sigma _z F\cos \omega _0 t - \sigma _z G(t).
\end{equation}
Here the first two terms describe an isolated TLS, which can model
a great variety of situations in physics and chemistry: from a
spin 1/2 particle in a magnetic field to superconducting flux and
charge qubits \cite{Makhlin}, \cite{Grif}. In order to be exact we
consider the first two terms in (\ref{Ham1}) to describe a
double-well system where only the ground states of the two wells
are occupied, with $\Delta$ being the energy splitting of a
symmetric ($\varepsilon=0$) TLS due to quantum tunnelling between
two wells. The quantity $\varepsilon$ is the bias, the external
energy parameter which makes the system asymmetric. The last two
terms in (\ref{Ham1}) describe the interaction with external
time-dependent high frequency, $F$, and low frequency, $G$ fields
which modulate the energy asymmetry between the two wells.

Hamiltonian ~(\ref{Ham1}) is written in the localized state basis, i.e., in the basis of states localized in each well. In terms of the
eigenstates basis, which we denote by upper case subscripts for the Pauli matrices $\sigma_X, \sigma_Y, \sigma_Z$, Hamiltonian ~(\ref{Ham1})
reads:
\begin{eqnarray}\label{Ham2}
H = \left[ {\frac{\Delta }{\Delta_{\varepsilon} }F\cos \omega _0 t
+
\frac{\Delta }{\Delta_{\varepsilon} }G(t)} \right]\sigma _X \nonumber\\
 +\left[{\frac{\Delta_{\varepsilon} }{2} - \frac{\varepsilon }{\Delta_{\varepsilon} }F\cos \omega _0 t
- \frac{\varepsilon }{\Delta_{\varepsilon} }G(t)} \right]\sigma
_Z,
\end{eqnarray}
where $\Delta_{\varepsilon} =\sqrt{\Delta^{2}+\varepsilon^{2}}$
is the gap between two energy states.

The inclusion of the dissipative environment in Hamiltonian (\ref{Ham1}), results in the Bloch-Redfield equations for the matrix operators
$\sigma_X$, $\sigma_Y$, $\sigma_Z$, \cite{Bloch}, \cite{Hartmann}. For weak driving ($F\ll\Delta$), and for weak coupling of the TLS to the bath
these equations can be approximated by Bloch-type equations~\cite{Saito}:

\begin{equation}\label{sigmaZ}
\left\langle {\dot \sigma _Z } \right\rangle  = \left( {2f\cos
\omega _0 t + 2g(t)} \right)\left\langle {\sigma _Y }
\right\rangle  - \Gamma _Z \left(\left\langle {\sigma _Z }
\right\rangle  - Z_0\right),
\end{equation}
\begin{align}\label{sigmaY}
\left\langle {\dot \sigma _Y } \right\rangle  =  - \left( {2f\cos
\omega _0 t + 2g(t)} \right)\left\langle {\sigma _Z }
\right\rangle \nonumber\\ + \left[ {\frac{\Delta_{\varepsilon}
}{\hbar } - \frac{{2\varepsilon }}{\Delta }f\cos \omega _0 t -
\frac{{2\varepsilon }}{\Delta }g(t)} \right]\left\langle {\sigma
_X } \right\rangle  - \Gamma \left\langle {\sigma _Y }
\right\rangle,
\end{align}
\begin{equation}\label{sigmaX}
    \left\langle {\dot \sigma _X } \right\rangle  =  - \left[ {\frac{\Delta_{\varepsilon} }{\hbar } -
    \frac{{2\varepsilon }}{\Delta }f\cos \omega _0 t - \frac{{2\varepsilon }}{\Delta }g(t)}
    \right]\left\langle {\sigma _Y } \right\rangle  - \Gamma \left\langle {\sigma _X }
    \right\rangle,
\end{equation}
where $f=\Delta$F$/\hbar\Delta_{\varepsilon}$, $g(t)=\Delta$G(t)$/\hbar\Delta_{\varepsilon}$, and
$Z_0=-\tanh\left(\Delta_\varepsilon/k_BT\right)$ is the equilibrium polarization of the system in the absence of external excitation sources
($f=0$, $g=0$). The angled brackets in Eqs.~(\ref{sigmaZ}), (\ref{sigmaY}), and (\ref{sigmaX}) denote the trace over reduced density matrix
$\rho(t)$, which is obtained by tracing out all environment degrees of freedom: $\langle\sigma_X\rangle=Tr(\sigma_X\rho(t))$, etc. In order to
simplify the problem we assume the  relaxation, $\Gamma_Z$ and dephasing, $\Gamma$, rates in (\ref{sigmaZ})-(\ref{sigmaX}) are time-independent,
i.e. the rates are slowly varying functions on the scale of Rabi period which is of the order of $1/f$.

Assuming that a high frequency driving amplitude $F$ is
sufficiently small we write the desired solution of
Eqs.~(\ref{sigmaZ}), (\ref{sigmaY}), and (\ref{sigmaX}) as:
\begin{equation}\label{sZ}
    \left\langle {\sigma _Z } \right\rangle  = Z(t),
\end{equation}
\begin{equation}\label{sY}
    \left\langle {\sigma _Y } \right\rangle  = Y(t) + A(t)\cos (\omega _0 t) + B(t)\sin (\omega _0 t),
\end{equation}
\begin{equation}\label{sX}
    \left\langle {\sigma _X } \right\rangle  = X(t) + C(t)\cos (\omega _0 t) + D(t)\sin (\omega _0 t),
\end{equation}
where  $X$, $Y$, $Z$, $A$, $B$, $C$, and $D$ are slowly varying, as compared to the high frequency $\omega_0$, quantities.

As is known, a two-level system resonantly irradiated with a high
frequency undergoes a low frequency Rabi oscillations. However, if
the external low frequency excitation, $G(t)$ is absent $(g=0)$,
the Rabi oscillations are damped out. For this case we obtain at
the degeneracy point ($\varepsilon=0$) the following solution:
$X=0$, $Y=0$, $C=B$, $D=-A$,
\begin{equation}\label{Zdamp}
    Z(t) = z^*  + z_0 e^{ -
\Gamma _1 t}  + e^{ - \Gamma _2 t} \left( {z_1 \cos \omega _R t +
z_2 \sin \omega _R t} \right),
\end{equation}
\begin{equation}\label{Adamp}
    A(t) = a^*  + a_0 e^{ - \Gamma _1
t}  + e^{ - \Gamma _2 t} \left( {a_1 \cos \omega _R t + a_2 \sin
\omega _R t} \right),
\end{equation}
\begin{equation}\label{Bdamp}
 B(t) = b^*  + b_0 e^{ - \Gamma _1 t}  +
e^{ - \Gamma _2 t} \left( {b_1 \cos \omega _R t + b_2 \sin \omega
_R t} \right),
\end{equation}
where the relaxation rates, $\Gamma_1$, $\Gamma_2$, and the Rabi
frequency $\omega_R$ are determined by the equation
\begin{equation}\label{lambda}
\left( {\lambda  - i\Gamma } \right)^2 \left( {\lambda  - i\Gamma _Z } \right) - \left( {\lambda  - i\Gamma } \right)f^2  - \left( {\lambda  -
i\Gamma _Z } \right)\delta ^2  = 0.
\end{equation}
Here $\lambda$ is the eigenvalue for the oscillation mode, e. g., $Z(t)\approx e^{i\lambda t}$. This cubic equation has simple analytic
solutions only for two cases. Firstly, in the absence of a relaxation in Eqs. (\ref{sigmaZ}) - (\ref{sigmaX}) $(\Gamma_Z=\Gamma=0)$ we obtain
$\Gamma_1=\Gamma_2=0$, $\omega_R=\sqrt{\delta^2+f^2}$, where $\delta$ is the high frequency detuning parameter,
$\delta=\omega_0-\Delta_\varepsilon/\hbar$.
 Secondly, for $\delta=0$ we obtain $\Gamma_1=\Gamma$, $\Gamma_2=(\Gamma+\Gamma_Z)/2$, $\omega_R=\sqrt{f^2-(\Gamma-\Gamma_Z)^2/4}$. For the
subsequent derivation we need only the quantities $z^*$, $a^*$, $b^*$ which can be determined as the steady state solution of the equations
(\ref{sigmaZ}), (\ref{sigmaY}), and (\ref{sigmaX}): $z^*=P_ZZ(0)$, $a^*=z^*f\Gamma/(\Gamma^2+\delta^2)$, $b^*=z^*f\delta/(\Gamma^2+\delta^2)$,
where $P_Z  = \frac{{\Gamma _Z \left( {\Gamma ^2  + \delta ^2 } \right)}}{{\Gamma _Z \left( {\Gamma ^2  +
    \delta ^2 } \right) + f^2 \Gamma }}$.

The quantity $P_ZZ_0$ is the nonequilibrium polarization, i. e., the steady state difference of occupation probabilities between two energy
levels in the case when the high frequency excitation is applied to the TLS. Therefore, in the absence of the low frequency excitation the Rabi
oscillations decay with a rate $\Gamma_2$ given by Eq. (\ref{lambda}). The main goal of our investigation is to obtain the persistent
oscillations of the quantities $Z(t)$ (\ref{sZ}), $Y(t)$ (\ref{sY}), $X(t)$ (\ref{sX}) which can be detected with low frequency (compared to the
gap) electronic circuitry. In what follows we show that a low frequency signal applied to TLS can sustain the persistent low frequency
oscillations of the above mentioned quantities. The amplitude of these oscillations has a clear resonance at the Rabi frequency with the width
of the resonance being dependent on the damping rates of the system.

The Eqs. (\ref{sigmaZ}), (\ref{sigmaY}), and (\ref{sigmaX}) are analized in the presence of a low frequency excitation $G(t)$. We insert
Eqs.~(\ref{sZ}), (\ref{sY}),  (\ref{sX}) into (\ref{sigmaZ}), (\ref{sigmaY}), and (\ref{sigmaX}), and in accordance with the ideology of
rotating wave approximation retain only the low frequency terms. We consider the force $G$ to be small, and therefore neglect the terms which
are of second and higher order in $G$. In addition, we keep only the terms which oscillate within the bandwidth of the Rabi frequency and
neglect the terms which are of the order of $\hbar f/\Delta_\varepsilon$, $\hbar\Gamma/\Delta_\varepsilon$, $\hbar \Gamma_Z/\Delta_\varepsilon$.
Consequently we obtain the following set of equations for low frequency quantities:
\begin{equation}\label{Z1}
    \dot Z = fA - \Gamma _ZZ
\end{equation}

\begin{equation}\label{Y1}
    \dot Y = \frac{\Delta_\varepsilon}{\hbar }X - \Gamma Y + 2gz^* -
    \frac{\varepsilon }{\Delta }fB
\end{equation}

\begin{equation}\label{X1}
    \dot X =  - \frac{\Delta_\varepsilon }{\hbar }Y - \Gamma X +
    \frac{\varepsilon }{\Delta }fA
\end{equation}
\begin{eqnarray}\label{A3}
\ddot A + 2\Gamma \dot A + (\Omega _R^2+\Gamma^2) A =
\nonumber\\f\left( {\Gamma _Z  - \Gamma } \right)Z -
 \Gamma \frac{{2\varepsilon }}{\Delta }gb^*- \delta \frac{{2\varepsilon }}
 {\Delta }ga^* -
 \frac{{2\varepsilon }}{\Delta }\dot gb^*
\end{eqnarray}

\begin{align}\label{B3}
 \ddot B + 2\Gamma \dot B + \left( {\delta^2 +\Gamma^2 } \right)B =\nonumber\\  - \delta fZ +
 \Gamma \frac{{2\varepsilon }}{\Delta }ga^* - \delta \frac{{2\varepsilon }}{\Delta }gb^* +
 \frac{{2\varepsilon }}{\Delta }\dot ga^*
 \end{align}
From these equations we can obtain the Fourier components for the
slowly varying quantities $A(\omega )$ and $B(\omega)$, and for
the low frequency persistent response of the TLS to a small low
frequency excitation, $\widetilde{Z}(\omega)$,
$\widetilde{Y}(\omega)$, and $\widetilde{X}(\omega)$:
$\widetilde{C}(\omega)=\widetilde{B}(\omega)$,
$\widetilde{D}(\omega)=-\widetilde{A}(\omega)$,
\begin{equation}\label{ZOmega}
\widetilde{Z}(\omega ) = \widetilde{g}(\omega )\frac{{2\varepsilon
}}{\Delta }f^2 P_Z Z_0\frac{\delta }{{\delta ^2  + \Gamma ^2
}}\frac{{2\Gamma  + i\omega }}{s(\omega)},
\end{equation}
\begin{equation}\label{AOmega}
\widetilde{A}(\omega ) = \widetilde{g}(\omega )\frac{{2\varepsilon
}}{\Delta }f{\kern 1pt} P_Z Z_0\frac{\delta }{{\delta ^2  +
    \Gamma ^2 }}
     \frac{{\left( {2\Gamma  + i\omega } \right)\left( {i\omega  + \Gamma _Z }
     \right)}}{s(\omega)},
\end{equation}

\begin{eqnarray}\label{BOmega}
 \widetilde{B}(\omega ) =  - \widetilde{g}(\omega )\frac{{2\varepsilon }}{\Delta }f{\kern 1pt} P_Z
 Z_0
 \frac{\delta }{{\delta ^2  + \Gamma ^2 }}
 \frac{1}{{\left( {\delta^2 +(\Gamma+i\omega)^2  } \right)}}
{\rm{}}\left\{ {\frac{{f^2 \delta \left( {2\Gamma + i\omega }
 \right)}}{s(\omega)} - \frac{{\delta ^2  - \Gamma ^2  - i\omega \Gamma }}{\delta }}
 \right\},
 \end{eqnarray}
\begin{equation}\label{YOmega}
\widetilde{Y}(\omega ) = \frac{\hbar }{\Delta_{\varepsilon}
}\frac{\varepsilon }{\Delta }f\widetilde{A}(\omega ) + \left(
{\frac{\hbar }{\Delta_{\varepsilon} }} \right)^2 \left(
{2\widetilde{g}(\omega )P_Z Z_0 - \frac{\varepsilon }{\Delta
}f\widetilde{B}(\omega )} \right)\left( {\Gamma  + i\omega }
\right),
\end{equation}
\begin{equation}\label{XOmega}
\widetilde{X}(\omega ) = \left( {\frac{\hbar
}{\Delta_{\varepsilon} }} \right)^2 \frac{\varepsilon }{\Delta
}f\widetilde{A}(\omega )\left( {\Gamma + i\omega } \right) -
\frac{\hbar }{\Delta_{\varepsilon} }\left( {2\widetilde{g}(\omega
)P_Z Z_0 - \frac{\varepsilon }{\Delta }f\widetilde{B}(\omega )}
\right),
\end{equation}

where   $s(\omega)=(\Omega _R-\omega +i\Gamma)(\Omega _R+\omega
-i\Gamma)(i\omega + \Gamma _Z)
       - f^2 \left( {\Gamma _Z  - \Gamma }
      \right)$; $\Omega_R=\sqrt{\delta^2+f^2}$ is the Rabi frequency in the
      absence of the damping ($\Gamma_Z=\gamma=0$). \footnote {In above
      expressions we disregard the Fourier components of the
      damping terms (Eqs. (\ref{Zdamp}), (\ref{Adamp}), (\ref{Bdamp})) since for the persistent signal $g(t)$ they are
unimportant.}

As can be concluded from these expressions the persistent low frequency oscillations
 of the spin components $Z(t)$, $X(t)$, $Y(t)$ appear as the
response to the low frequency external force, $g(t)$, only in the
presence of the high frequency excitation ($f\neq0$). Exactly at
resonance ($\delta=0$) $\widetilde{A}(\omega)=0$, and
$\widetilde{Z}(\omega)=0$, but $\widetilde{B}(\omega)\neq0$. At
this point the population of the two levels are equalized, and
spin circularly rotates in the $XY$ plane with frequency
$\omega_0$, with the center of the circle being precessed with the
frequency $\omega$ of the low frequency external source.

\begin{figure}[tbp]
\centerline{\includegraphics[width=8cm]{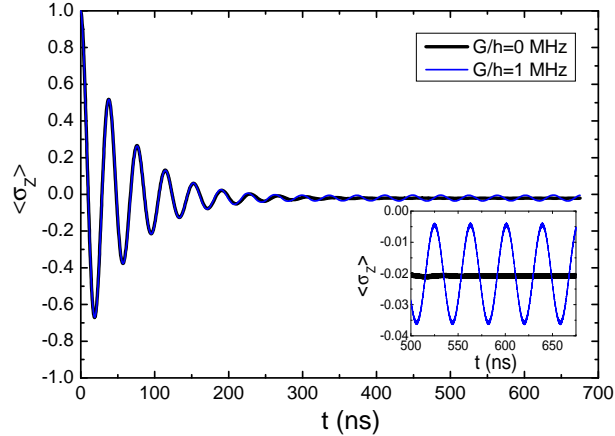}} \caption{Time evolution of $<\sigma_z>$.  (thick) G=0, (thin) $G/h=1$~MHz. The insert shows
the undamped oscillations of $<\sigma_z>$ at $G/h=1$~MHz.} \label{fig1}
\end{figure}

 As an example we show below the time evolution of the
quantity $\langle\sigma_Z\rangle(t)=Z(t)$, obtained from the
numerical solution of the equations (\ref{sigmaZ}),
(\ref{sigmaY}), and (\ref{sigmaX}) where we take a low frequency
excitations as $G(t)=Gcos(\omega_L t)$. The calculations have been
performed with initial conditions $\langle\sigma_Z\rangle(0)=1$,
$\langle\sigma_X\rangle(0)$=$\langle\sigma_Y\rangle(0)=0$ for the
following set of the parameters: $F/h=36$~MHz, $\Delta/h=1$~GHz,
$\Gamma/2\pi=4$~MHz, $\Gamma_z/2\pi=1$~MHz, $\epsilon/\Delta=1$,
$Z_0=-1$, $\delta/2\pi=6.366$~MHz, $\omega_L/\Omega_R=1$. As is
seen from Fig.\ref{fig1} in the absence of low frequency signal
($G=0$) the oscillations are damped out, while if $G\neq0$ the
oscillations persist.

\begin{figure}[tbp]
\centerline{\includegraphics[width=8cm]{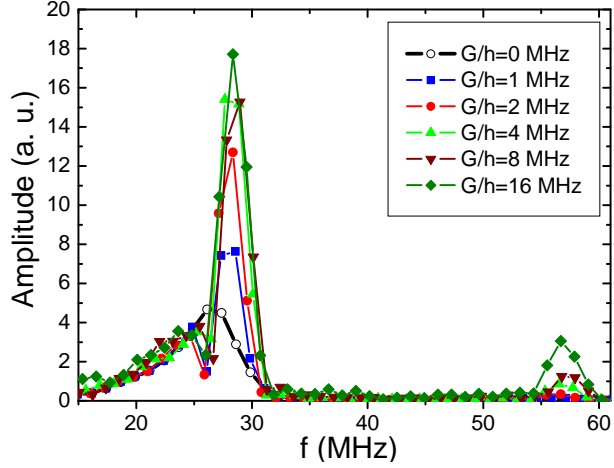}}
\caption{Fast Fourier transform of $<\sigma_z>$ at different amplitudes $G/h$ of
low-frequency field.} \label{fig2}
\end{figure}

\begin{figure}[tbp]
\centerline{\includegraphics[width=8cm]{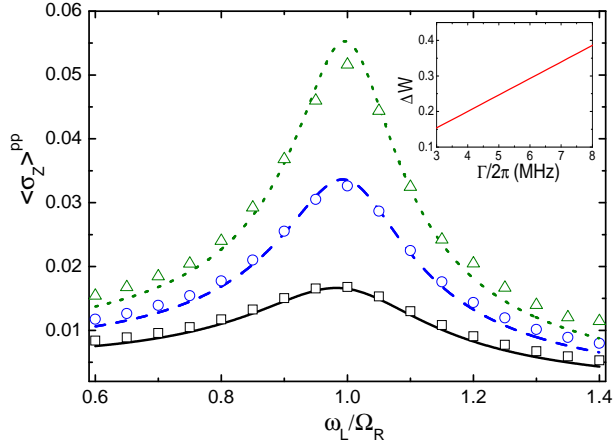}}
\caption{Peak-to-peak amplitude of $<\sigma_z>^{pp}$ undamped
harmonic signal at $t\approx700$~ns vs $\omega_L/\Omega_R$.
$G/h=1$~MHz, $\Gamma_Z/2\pi=1$~MHz, (solid curve)
$\Gamma/2\pi=6$~MHz, (dashed curve) $\Gamma/2\pi=4$~MHz, (dotted
curve) $\Gamma/2\pi=3$~MHz. The curves and the width of resonance,
$\Delta W$ at $1/\sqrt{2}$ level (see the insert) are calculated
from Eq.~(\ref{ZOmega}); the symbols are found from numerical
solution of Eqs.~(\ref{sigmaZ}),(\ref{sigmaY}),(\ref{sigmaX}).}
\label{fig3}
\end{figure}

The Fourier spectra of these signals are shown on Fig.\ref{fig2} for different amplitudes of low frequency excitation. For $G=0$ the Rabi
frequency is positioned at approximately 26.2 MHz, which is close to $\Omega_R=26.24$ MHz. With the increase of $G$ the peak becomes higher. It
is worth noting the appearance of the peak at the second harmonic of Rabi frequency. This peak is due to the contribution of the terms on the
order of $G^2$ which we omitted in our theoretical analysis.
The comparison of analytical and numerical resonance curves
calculated for low frequency amplitude, $G/h=1$ MHz and different
dephasing rates, $\Gamma$ are shown on Fig.\ref{fig3}. The curves
at the figure are the peak-to-peak amplitudes of oscillations of
$Z(t)$ calculated from Eq.~(\ref{ZOmega}) with
$\widetilde{g}(\omega)=g(\delta(\omega+\omega_L)+\delta(\omega-\omega_L))/2$,
where $\delta(\omega)$ is Dirac delta function. The point symbols
are found from numerical solution of
Eqs.~(\ref{sigmaZ}),(\ref{sigmaY}),(\ref{sigmaX}). The widths of
the curves depend on $\Gamma$ (see the insert) and the positions
of the resonances coincide with the Rabi frequency. A good
agreement between numerics and Eq.~\ref{ZOmega}, as shown at
Fig.~\ref{fig3}, is observed only for relative small low frequency
amplitude $G/h$, for which our linear response theory is valid.

The key point of the method we described above is that it allows for the detection of the high frequency response of a TLS at a frequency which
is much less than the gap frequency. The low frequency dynamics of the quantities  $\left\langle {\sigma _X } \right\rangle, \left\langle
{\sigma _Y } \right\rangle$, $\left\langle {\sigma _Z } \right\rangle$ bears the information about the relaxation, $\Gamma_Z$, and dephasing,
$\Gamma$, rates.

The experimental realization of this method (the detection scheme)
depends on the problem under investigation. For example, the
method can easily be adapted for NMR. In this case the quantities
$\left\langle {\sigma _Z } \right\rangle, \left\langle {\sigma _X
} \right\rangle$, $\left\langle {\sigma _Y } \right\rangle$ are
the longitudinal, $M_Z$, and the transversal polarizations of the
sample, $M_X$, and $M_Y$. Indeed, from the comparison of the
Hamiltonian ~(\ref{Ham2}) and the equations of motion
(\ref{sigmaZ}), (\ref{sigmaY}), (\ref{sigmaX}) with those for
nuclear spin ($H=-\vec{\mu} \vec{B}$, $d\vec{\mu}/dt=\gamma
[\vec{\mu} \vec{B}]$, where $\gamma$ is the gyromagnetic ratio),
it is clear that the NMR case corresponds to the polarization of a
sample with a field $B_0$ along the $z$ axes with a high frequency
excitation $B_1cos(\omega_{0}t)$ and a low frequency probe $G(t)$
being applied in the ZX plane at an angle $\theta$ to the $z$
axis: $B_Z=B_0+cos\theta (B_1cos(\omega_{0}t)+G(t))$,
$B_X=-sin\theta (B_1cos(\omega_{0}t)+G(t))$, $B_Y=0$. This analogy
allows for the direct application of Eqs.
(\ref{ZOmega})-(\ref{XOmega}) to the nuclear spin with the
following substitutions: $\Delta_\varepsilon=\gamma\hbar B_0$,
$\varepsilon/\Delta_\varepsilon=cos\theta$,
$\Delta/\Delta_\varepsilon=sin\theta$, $f=\gamma B_1sin\theta$,
$g(\omega)=\gamma G(\omega)sin\theta$,
$\varepsilon/\Delta=\cot\theta$, $Z_0=M_0$ - the equilibrium
magnetization of a sample. In NMR all low frequency components of
the magnetization, $\widetilde{Z}(\omega)$ (\ref{ZOmega}),
$\widetilde{Y}(\omega)$ (\ref{YOmega}), $\widetilde{X}(\omega)$
(\ref{XOmega}), and their combinations are accessible for the
measurements.

Our method can be directly applied to a persistent current qubit,
which is a superconducting loop interrupted by three Josephson
junctions ~\cite{Mooij}, \cite{Orlando}. For these qubits the
successful experimental implementation of low frequency readout
electronics has been demonstrated ~\cite{Greenberg3, Grajcar,
Il'ichev}. The average current in the qubit loop is proportional
to the low frequency part of the quantity $\langle {\sigma
_Z}\rangle(t) $, which is directly connected to the probabilities
of occupation of the ground, $P_-(t)$, and the excited, $P_+(t)$
states: $\langle {\sigma _Z}\rangle(t)$=$P_-(t)$-$P_+(t)$.
Therefore, this current can be detected through the variation of
its magnetic flux either by a DC SQUID ~\cite{Lupascu} or by a
high quality resonant tank circuit inductively coupled to the
qubit ~\cite{Greenberg3, Grajcar, Il'ichev}.

In conclusion, we proposed a method to study the Rabi oscillations
in a dissipative TLS by irradiating it simultaneously with high,
resonant, and low frequency. The low frequency response of a
system has a clear resonance at the Rabi frequency with the
resonance width being dependent on the damping rates. Therefore,
the method allows for the experimental determination, in the low
frequency domain, of the relaxation and dephasing rates of
dissipative two-level systems.

\begin{acknowledgments}
We are grateful to D. Averin, M. Grajcar, A. Korotkov, A.
Shnirman, A. Smirnov, A. Zagoskin, A. Maassen van den Brink, W.
Krech and V. Shnyrkov for fruitful discussions.

The authors acknowledge the support from D-Wave Systems. Ya.~G.
acknowledges partial support by the INTAS grant 2001-0809. E.~I.
thanks the EU for support through the RSFQubit project.

\end{acknowledgments}

\end{document}